\documentclass{appolb}
\usepackage{epsfig,pstricks}
\usepackage{graphicx}
\newcommand{\lsim}{\stackrel{\scriptstyle <}{\phantom{}_{\sim}}}
\newcommand{\gsim}{\stackrel{\scriptstyle >}{\phantom{}_{\sim}}}

\begin{document}
\pagestyle{plain}
\title{Cooling of neutron stars with stiff stellar matter%
\thanks{Presented at the International Conference on Critical Point and Onset of Deconfinement (CPOD'2016), May, 29 -- June, 5, 2016, University of Wroclaw}
\\
\author{H.~Grigorian$^{1,2}$, D.~N.~Voskresensky$^3$, D.~Blaschke$^{3,4,5}$\\
\address{
	\begin{center}
$^1$~Laboratory of Information Technologies, JINR, RU-141980 Dubna\\
$^2$~Department of Theoretical Physics, Yerevan State University, AM-0025 Yerevan\\
$^3$~National Research Nuclear University (MEPhI), RU-115409 Moscow\\
$^4$~Institute of Theoretical Physics, Wroclaw University, PL-50-204 Wroclaw\\
$^5$~Bogoliubov Laboratory of Theoretical Physics, JINR, RU-141980 Dubna
\end{center}
}}}
\maketitle
\begin{abstract}
Recent evidence for 
high masses ($\sim 2~ M_\odot$)  pulsars PSR J1614-2230 and PSR J0348-0432 requires neutron
star matter to have a  stiff equation of state (EoS).
The thermal evolution of compact stars (CS) with stiff hadronic EoS necessitates the application of the ``nuclear medium cooling" scenario with a selection of appropriate proton gap profiles together with  in-medium effects (like pion softening) on cooling mechanisms in order to achieve a satisfactory explanation of all existing observational data for the temperature-age relation of CS. Here we focus on two examples from  \cite{Grigorian:2016leu} for a stiff hadronic EoS  without (DD2 EoS) and with (DD2vex) excluded volume correction.
\end{abstract}

\section{Introduction}
The neutrino emissivities of various processes taking place in deeper layers of compact stellar objects are depending on the behavior of the nucleon-nucleon ($NN$) interaction and  the proton
and neutron paring gaps as functions of the density.
 An information about these cooling processes,  the heat transport and the EoS of  compact star (CS) matter can be gained from analyzing the  data on the evolution of surface temperatures of CS.

The available data (see Fig.~\ref{Fig:Cool2} below) can be separated into three groups related to slow cooling objects (labeled 8; 5; 1; 2; 4; A in Fig.~\ref{Fig:Cool2}), intermediate cooling (3; 6; 7; Cas A; B; E) and rapid cooling (C; D) objects.
In order to explain the difference in the cooling of the slowly and rapidly cooling objects a three order of magnitude difference in their luminosities is required.
Therefore, one needs to take into account the strong density-dependence (and thus neutron star (NS) mass dependence) of the medium modification of the $NN$ interaction.  Otherwise it is not easy to appropriately explain the essentially different surface temperatures of various objects in the hadronic scenario within the so called ``minimal cooling paradigm'', cf.
\cite{Elshamouty:2013nfa}, where the only relevant rapid process is the so called pair-breaking-formation (PBF) process on neutrons paired in the $3P_2$ channel.

The main medium modification is caused by the softening of the pion exchange contribution with increasing density for $n\gsim n_0$, where $n_0\simeq 0.16$ fm$^{-3}$ is the nuclear saturation density, and by the density dependent superfluid pairing gaps, see \cite{Migdal:1990vm,Voskresensky:2001fd} for details.
So the cooling of various sources should be essentially different due to the difference in their masses. This effect of course strongly depends on the density profiles of star structure and thus on the EoS.
The recent measurements of the high masses of the pulsars PSR J1614-2230
\cite{Demorest,Fonseca:2016tux} and PSR J0348-0432 \cite{Anotoniadis} on the one hand and of the low masses for PSR J0737-3039B \cite{Kramer} for the companion of PSR J1756-2251 \cite{Faulkner,Ferdman:2014rna} on the other have provided the proof for the existence of CS with masses varying  at least from 1.2 to 2.0 $M_\odot$.
 It requires a relatively stiff EoS to reach the high value of the mass on the stable branch of CS.
On the other hand   for a stiff EoS   lower densities are reached in the cores of stars with the same mass compared to those for a soft EoS.

\section{EoS}

\begin{figure}[!thb]
 \centering
 	\includegraphics[width=0.8\textwidth]{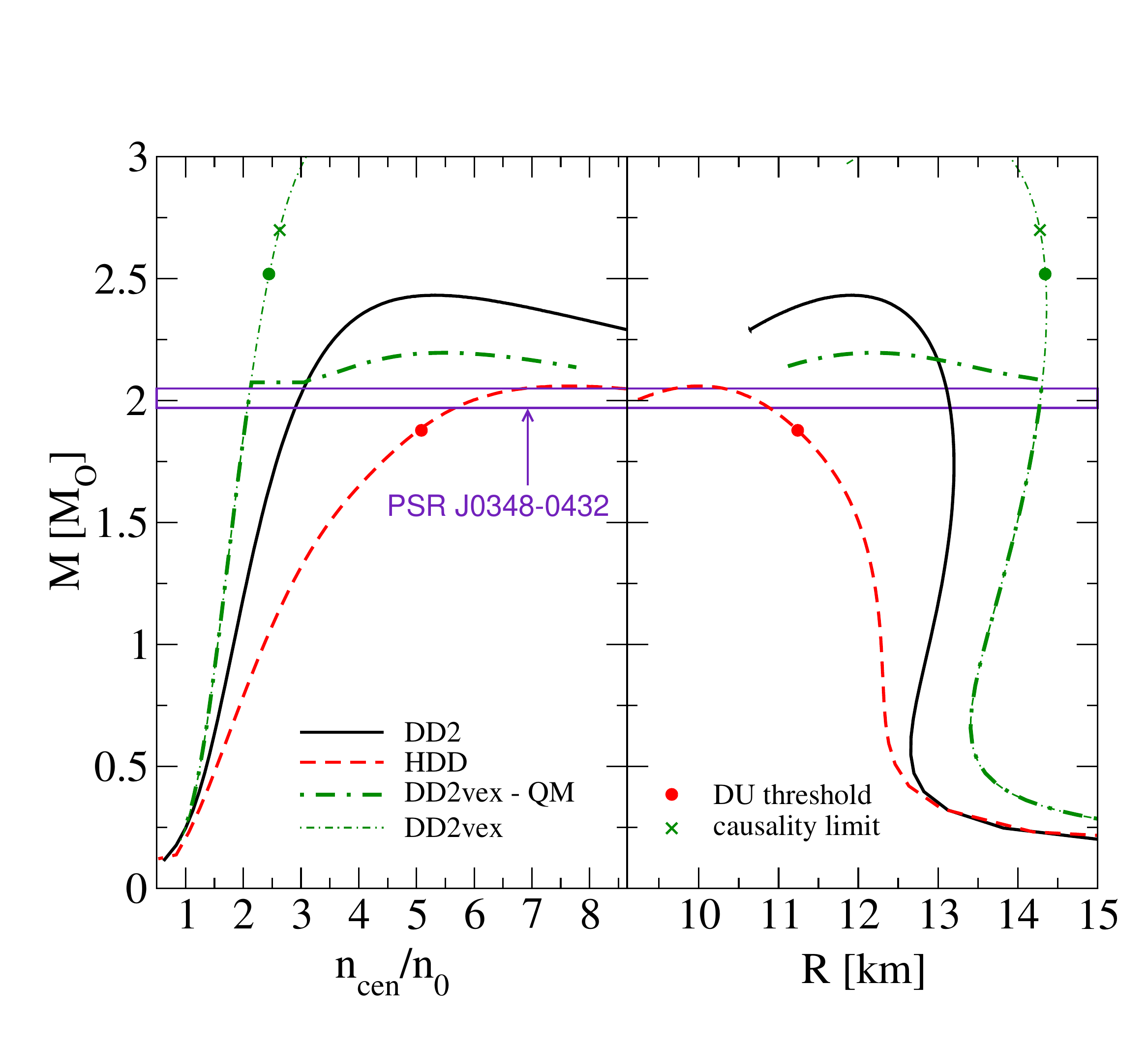}
	\caption{Mass vs. central baryon density (left) and  radius
 		(right) for the HDD  (dash lines), a stiffer DD2  \cite{Typel:2009sy} (solid lines) and a still stiffer DD2vex hadronic EoS (dash-dotted lines). Bold dash-dotted continuations show a possible quark (QM) transition. The crosses for the DD2vex hadronic EoS indicate violation of the causality. The band shows the  data on the measured mass for the pulsar PSR J0348-0432. The bold dots indicate the DU thresholds on hadronic branches. }
 	\label{Fig:MR}
\end{figure}

The HHJ hadronic EoS, which has been exploited in our previous works, was stiffened  in \cite{Blaschke:2013vma} for $n> 4n_0$ to comply with the constraint that the EoS
should allow for a maximum NS mass above the value $M=2.01\pm 0.04~M_\odot$ measured for
PSR J0348+0432 by \cite{Anotoniadis}, see also \cite{Demorest,Fonseca:2016tux}.
However, the resulting EoS (labeled as HDD), which produces
$M_{\rm max}=2.06~M_{\odot}$,  still might be not sufficiently stiff, since the existence of even more massive objects than
those known up to now \cite{Demorest,Fonseca:2016tux,Anotoniadis} is not excluded. In Fig. \ref{Fig:MR} we show the mass - central density and mass - radius relationship for different EoS models used in our simulations of cooling evolution of CS.

\section{Effective pion gap and $pp$ pairing gaps}

\begin{figure}[!thb]
	\centering
$	\begin{array}{cc}
	\includegraphics[width=0.5\textwidth]{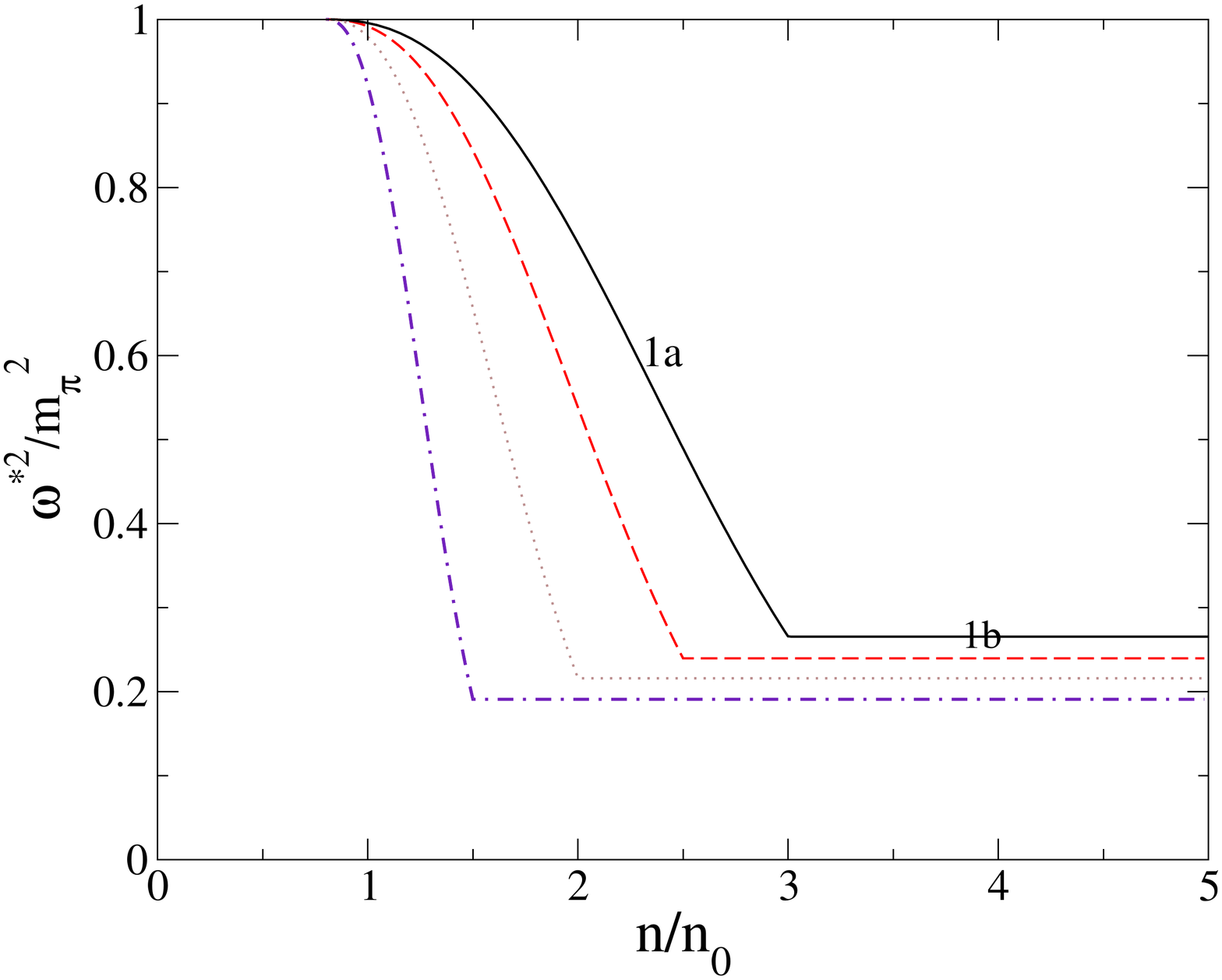}&
	\includegraphics[width=0.5\textwidth]{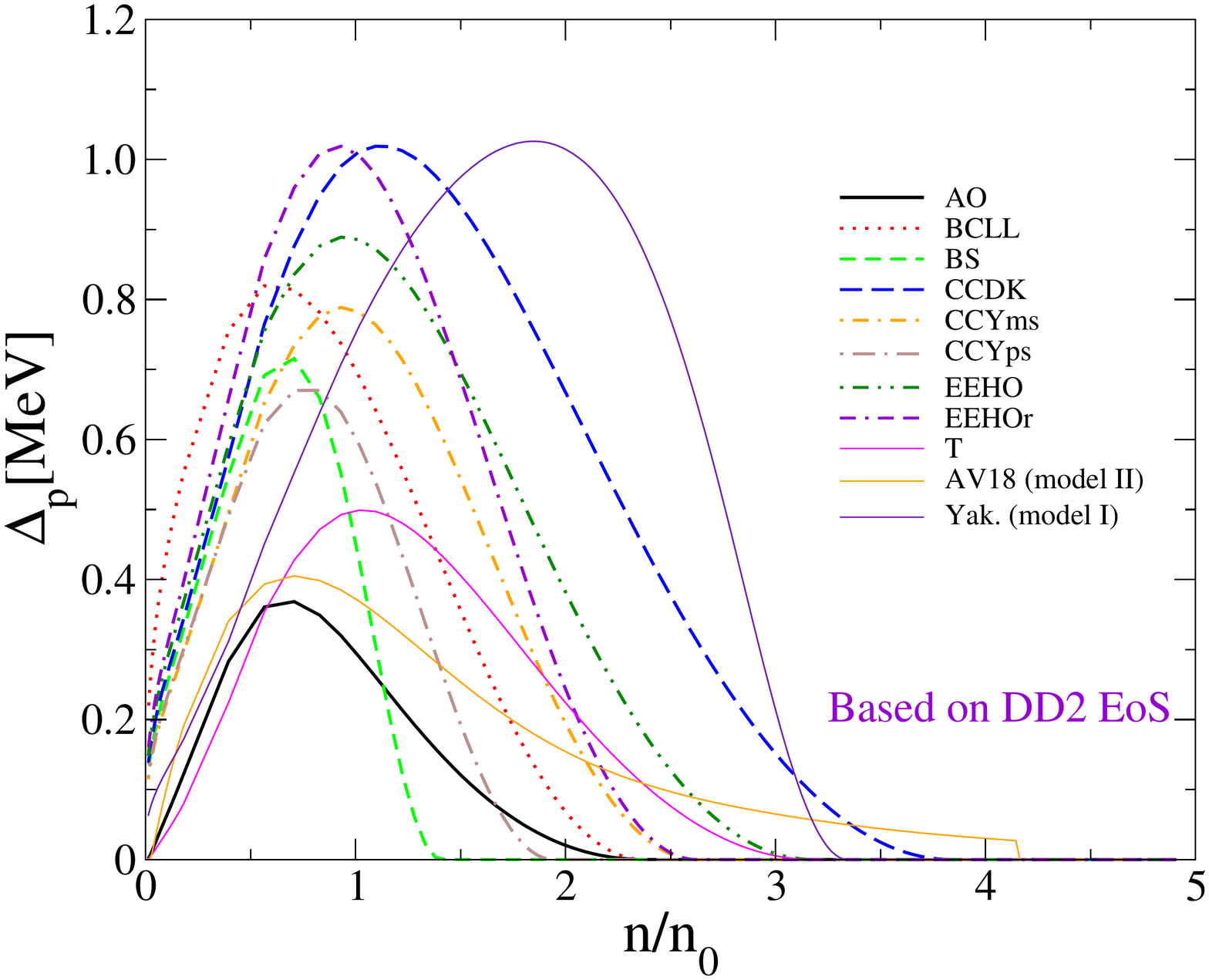}
		\end{array}$
	\caption{Left panel: Square of the effective pion gap as a function of the density  (curves labeled a),  with a pion softening saturated above a critical value $n_c^\pi$  (i.e., disregarding a possibility of a pion
		condensation), see lines labeled b, $m_{\pi}$ is the pion mass. The solid line (1a+1b), $n_c^\pi =3n_0$,  corresponds to
		the same parametrization as in  \cite{Blaschke:2004vq}, the other lines, $n_c^\pi =2.5 n_0$, $2n_0$ and $1.5 n_0$ demonstrate a stronger pion  softening effect. Right panel:
$1S_0$ $pp$ gaps as  functions of the baryon  density for zero temperature.
		The abbreviations in the legend correspond to those used in Ref.~\cite{Ho:2014pta}.
		The gaps labeled as ``Yak" and ``AV18" are those (models I and II, respectively)
		exploited in our previous works
		\cite{Blaschke:2004vq,Grigorian:2005fn,Blaschke:2011gc,Blaschke:2013vma} (right panel).}
	\label{Fig:omegatilde}
\end{figure}

The most efficient processes within the hadronic ``nuclear medium cooling" scenario, cf. \cite{Voskresensky:2001fd}, are the medium modified Urca (MMU) processes, like   $nn\to npe\bar{\nu}$,  i.e. the modified Urca (MU) processes computed by taking into account pion softening effects \cite{Migdal:1990vm} (see Fig.~\ref{Fig:omegatilde} left), and the
PBF processes $N\to N_{\rm pair}\nu\bar{\nu}$, $N=n$ or $p$.
While being enhanced owing to their one-nucleon nature
the latter processes are allowed only in the presence of nucleon pairing and should be computed by taking into account  in-medium effects in the weak interaction vertices. Various parametrizations for the effective pion gap being exploited here are shown in Fig.~\ref{Fig:omegatilde} left.

Note that in the minimal cooling scenario one assumes that the emissivity of the PBF process on protons is suppressed by two orders of magnitude compared to that for the PBF process on neutrons since the authors use the free $p\to p\nu\bar{\nu}$ vertices, whereas in matter the  decay may occur through the neutron and neutron hole and the electron and electron hole in the intermediate states of the reaction, cf. \cite{Voskresensky:2001fd}.
Note \cite{Migdal:1990vm} that the contribution of the intermediate reaction states is the largest in the emissivity of the MMU processes for $n\gsim n_0$. This contribution is not incorporated at all in the MU emissivity used in the minimal cooling scheme.

The resulting cooling curves are rather insensitive to the values of the $1S_0$ $nn$ pairing gaps but sensitive to the choice of the $1S_0$ $pp$ gaps and $3P_2$ $nn$ gaps, since $1S_0$ $nn$ pairing gaps drop already for $n\gsim 0.6 - 0.8 n_0$, whereas $1S_0$ $pp$ gaps  are spreading  to $n\sim 1.5 - 4 n_0$, and $3P_2$ $nn$ gaps may spread to a higher density, see Fig. \ref{Fig:omegatilde} right panel.
Indeed,  the dense interior rather than the crust determines the total luminosity within our scenario.
The gaps are sensitive to in-medium ($NN$ loop) effects, and their values are badly known due to  exponential dependence on the $NN$ interaction amplitude  in the pairing channel.
Especially the values of the $3P_2$ neutron pairing gaps are poorly known.
Ref.~\cite{Schwenk:2003bc}  computed a tiny value of $\Delta({ 3P_2}) \lsim 10$ keV.
 In our model an overall fit of the CS cooling data is obtained for a strongly suppressed value of the $3P_2$ neutron pairing gap, thus being in favour of this results.
The dependence of the cooling curves on the $3P_2$ neutron  and $1S_0$ proton paring gaps was studied within our scenario in  \cite{Grigorian:2005fn}.
The successful description of all cooling data within our scenario, where many  in-medium effects are shown to be important while they are being disregarded in the minimal cooling scenario, demonstrates
that the statement made within the latter scenario in some works, that an appropriate fit of existing  Cas A data  allows  to ``measure" the critical temperature of the $3P_2$ $nn$ pairing as $T_c \sim (5 - 9) \cdot 10^8$ K, is not justified. Note that the authors of the recent paper \cite{Ho:2014pta}
have explained the Cas A ACIS-S data for the NS mass $M=1.44~M_{\odot}$
within the minimal cooling scenario using the stiff BSk21 EoS,  a large proton gap and a moderate $3P_2$ neutron gap.
Hottest and coldest objects, however, can hardly be explained appropriately within the same scenario.

\begin{figure}[!thb]
\vspace{-1cm}
	\hspace{-5mm}
 	\includegraphics[width=0.6\textwidth,angle=0]{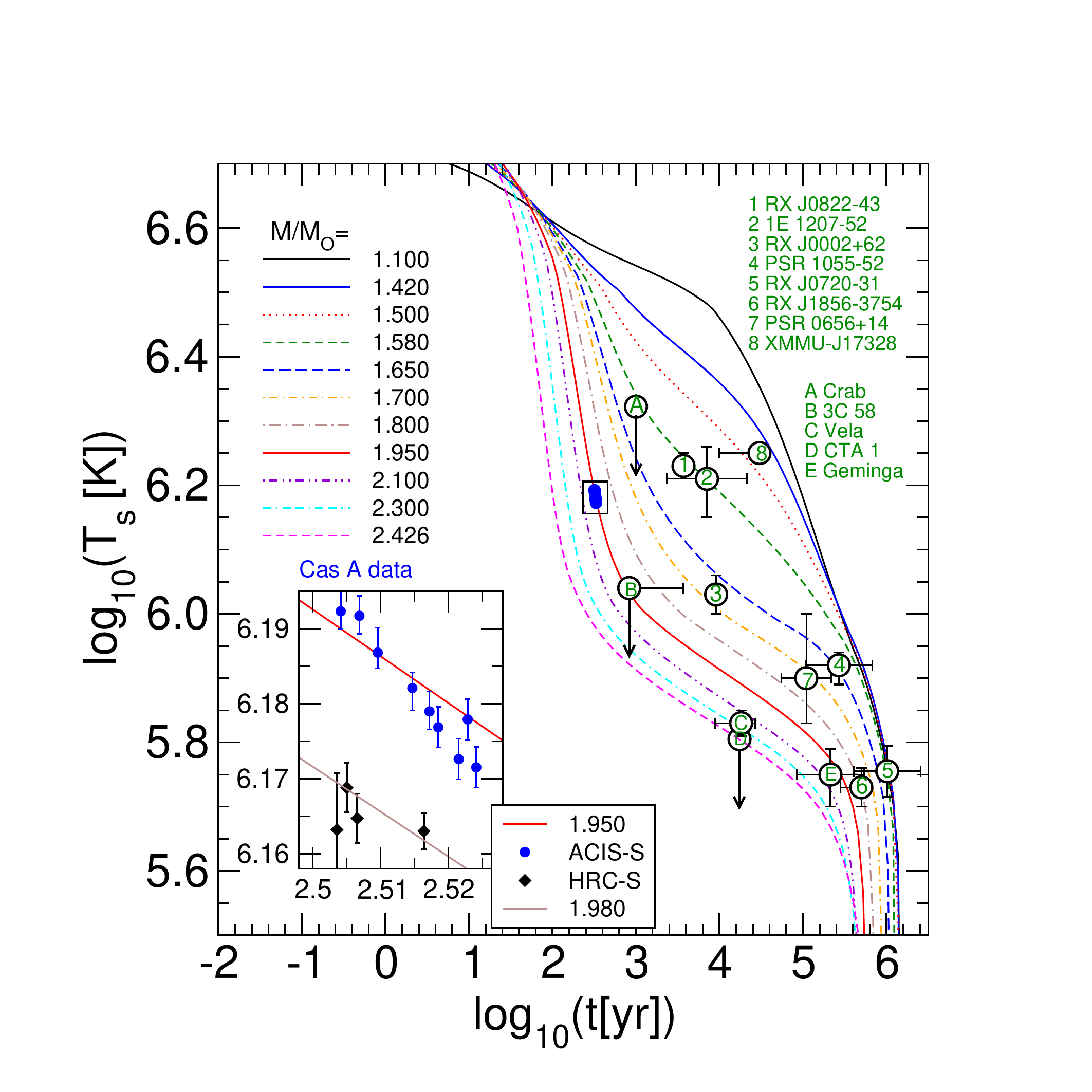}
	\hspace{-1cm}
 	\includegraphics[width=0.6\textwidth,angle=0]{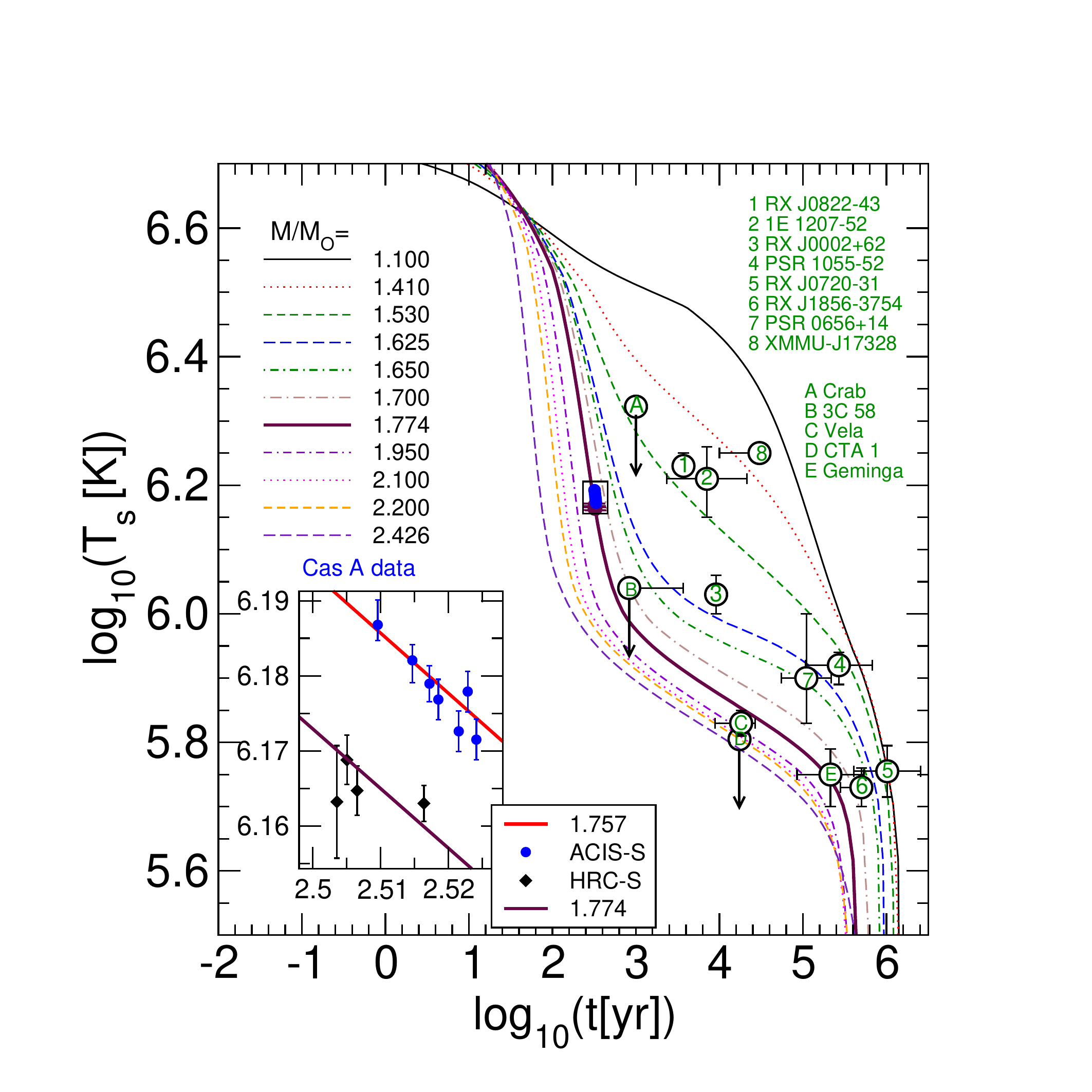}
 	\caption{Cooling curves for a NS sequence according to the  hadronic DD2 EoS; $n_c^\pi =3 n_0$, EEHOr (left) and $n_c^\pi =2.5 n_0$, CCYms  models (right),  see the text.
 	}
 	\label{Fig:Cool2}
 \end{figure}

      \begin{figure}[!htb]
\vspace{-1cm}

	\hspace{-5mm}
     	\includegraphics[width=0.6\textwidth]{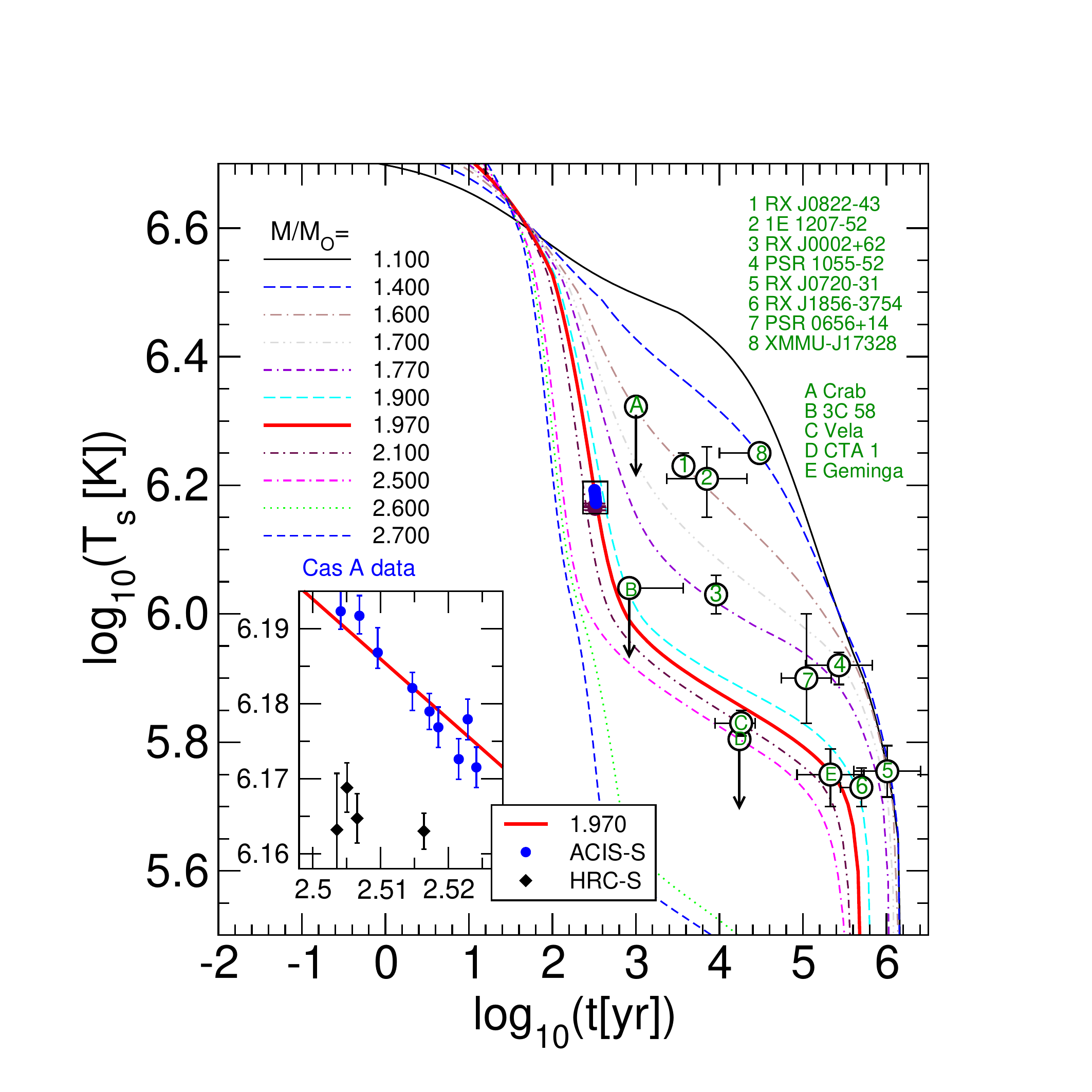}
	\hspace{-1cm}
      	\includegraphics[width=0.6\textwidth]{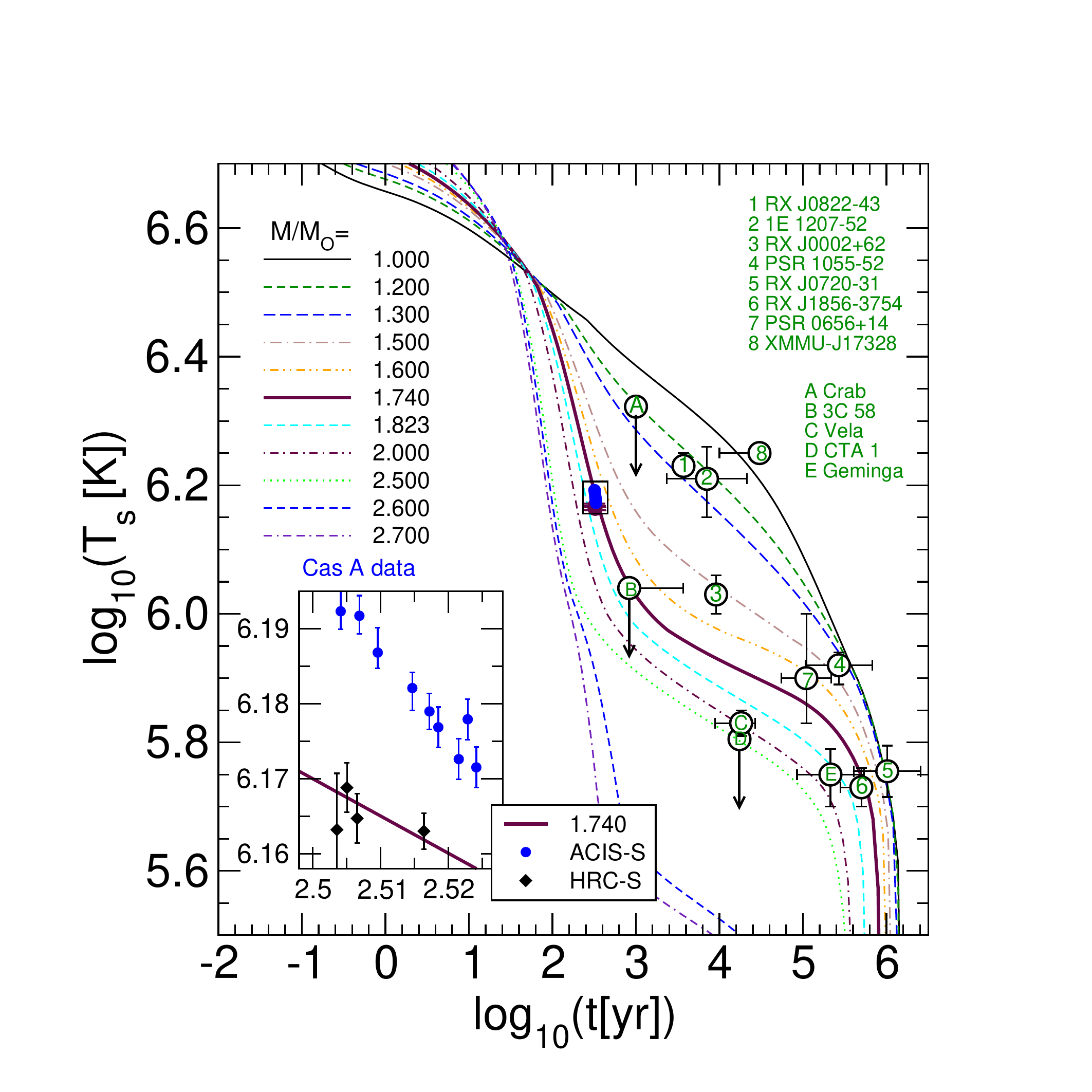}
	\caption{
     		      Cooling curves for a NS sequence according to the  hadronic
     		      DD2vex  EoS; $n_c^\pi =2.5 n_0$, BCLL (left) and AO (right) $pp$ pairing models, for details see the text.}
     	\label{Fig:Cool8b}
     \end{figure}

\section{Cooling}


In Fig.~\ref{Fig:Cool2} we demonstrate the CS  surface temperature $T_s$ and age $t$ relation. 
In the left panel we present results for
the effective pion gap  given by the solid curve 1a+1b in Fig.~\ref{Fig:omegatilde} left, $n_c^{\pi}=3~n_0$.
The $1S_0$ $pp$ pairing gap corresponds to model EEHOr. In the right panel we exploit the dashed curve of Fig.~\ref{Fig:omegatilde} left, $n_c^{\pi}=2.5~n_0$, and  $1S_0$ $pp$ pairing gap model CCYms (cf. Fig.~\ref{Fig:omegatilde} right panel).
The  mass range is shown in the legend.
Comparison with Cas~A ACIS-S and HRC-S data is shown in the inset.
The cooling data for Cas A from the ACIS-S instrument are explained by a NS mass $M=1.950~M_\odot$ (those of the HRC-S instrument by $M=1.980~M_\odot$), and $M=1.757~M_{\odot}$ (and $M=1.774~M_{\odot}$) correspondingly.

The CS  surface temperature $T_s$ vs. age $t$  for the model EoS DD2vex is shown in the 
Fig.~\ref{Fig:Cool8b} left and right panels. 
The effective pion gap corresponds to  $n_c^{\pi}=2 n_0$,  the dotted curve a+b in Fig.~\ref{Fig:omegatilde} left. 
In the left panel the $1S_0$ $pp$ pairing gap corresponds to the model BCLL  (Fig.~\ref{Fig:omegatilde} right panel). 
In the right panel the $1S_0$ $pp$ pairing gap corresponds to the  model AO (Fig.~\ref{Fig:omegatilde} right panel). 
Cas~A cooling data from the ACIS-S  and HRC-S instruments are explained with a NS of 
$M=1.970~M_\odot$ and  $M=1.740~M_\odot$, correspondingly.

      \section{Conclusions}
If in the future a very massive compact star (CS), and/or a CS of a middle mass but with a large radius would be observed, this would provide arguments for a stiff equation of state (EoS), like DD2 or even DD2vex, see Fig.~\ref{Fig:MR}.
Both EoS that we used, DD2 and DD2vex,  are compatible with the existing CS cooling data provided we exploit the nuclear medium cooling scenario developed in our previous works, under the assumption that different sources have different masses.
The resulting cooling curves prove to be  sensitive to the value and the density dependence of the $pp$ pairing gap and the effective pion gap. Both ACIS-S and HRC-S data for Cas~A can be fitted with the help of a variation of the model parameters. A stiffer EoS suggests a stronger pion softening effect.

Another consequence of stiff hadronic EoS which was not elaborated in this contribution is a phase transition to deconfined quark matter at low densities but high mass, as shown in Fig.~\ref{Fig:MR}.  
If the stable hybrid star branch of such models could be populated in nature, it would add another strip of 
cooling curves in the temperature-age diagram, either separated from the hadronic cooling curves discussed here or overlapping with them, depending on the characteristics of the cooling processes in quark matter. 
In this context it is interesting to discuss the cooling of high-mass twin stars 
\cite{Benic:2014jia,Alvarez-Castillo:2016oln} which, if they would exist in nature, populate a separate branch of stable hybrid stars, disconnected from the hadronic one.  
The discovery of these objects would imply the existence of a critical endpoint in the QCD phase diagram \cite{Blaschke:2013ana,Alvarez-Castillo:2016wqj}.
The investigation of their role for CS cooling scenarios is an important task which we take up in subsequent work.

\section*{Acknowledgments}

This work was supported by the NCN Opus programme under contract UMO-2014/13/B/ST9/02621 and by the Ministry of
Education and Science of the Russian Federation (Basic part).
We acknowledge the support from the MEPhI Academic Excellence Project under contract number 02.a03.21.0005 and from the Bogoliubov--Infeld and Ter-Antonian--Smorodinsky programmes as well as from the COST Action MP1304 "NewCompStar".

\end{document}